# Tunable two-dimensional Dirac nodal nets

Ding-Fu Shao,[1,*,‡] Shu-Hui Zhang,[2,‡] Xiaoqian Dang,[1,‡] and Evgeny Y. Tsymbal[1,†]

[1] *Department of Physics and Astronomy & Nebraska Center for Materials and Nanoscience,*
*University of Nebraska, Lincoln, Nebraska 68588-0299, USA*

[2] *College of Science, Beijing University of Chemical Technology,*
*Beijing, 100029, People's Republic of China*

Nodal line semimetals are characterized by symmetry-protected band crossing lines and are expected to exhibit nontrivial electronic properties. Connections of the multiple nodal lines, resulting in nodal nets, chains, or links, are envisioned to produce even more exotic quantum states. In this work, we propose a feasible approach to realize *tunable* nodal line connections in real materials. We show that certain space group symmetries support the coexistence of the planar symmetry enforced and accidental nodal lines, which are robust to spin-orbit coupling and can be tailored into intricate patterns by chemical substitution, pressure, or strain. Based on first-principles calculations, we identify non-symmorphic centrosymmetric quasi-one-dimensional compounds, $K_2SnBi$ and $MX_3$ (M = Ti, Zr, Hf and X = Cl, Br, I), as materials hosting such tunable 2D Dirac nodal nets. Unique Landau levels are predicted for the nodal line semimetals with the 2D Dirac nodal nets. Our results provide a viable approach for realize the novel physics of the nodal line connections in practice.

Quantum materials have recently become a promising platform for the discovery of new fermionic particles and novel quantum phenomena [1]. Among them are Weyl and Dirac semimetals, the three-dimensional (3D) materials with nontrivial band crossings at discrete points in the momentum space [2,3]. These materials support the quasiparticles resembling the relativistic Dirac (four-fold degenerate) and Weyl (doubly degenerate) fermions known from high-energy physics [2-10]. It has been demonstrated that the Dirac or Weyl cones can be tilted [11-13], the low-energy dispersion can be quadratic or cubic [14, 15], and the fermionic quasiparticles can hold three-, six-, or eight-fold degeneracies [3, 16-19], which do not have analogy in high-energy physics.

In addition to the nodal points, band crossings can also occur along the nodal lines [2, 20-26], resulting in unusual surface states and magneto-transport properties [21, 27, 28]. Importantly, the nodal lines can serve as the constituents for other nontrivial states. For example, the multiple nodal lines can form nodal chains [29-32], nets [33, 34], and links [35-38]. Nodal lines intersections can produce triple or four-fold degenerate points in non-centrosymmetric materials [29, 39] and support photoinduced Floquet multi-Weyl fermions [40, 41].

All the current studies of the interconnecting nodal line systems with non-negligible spin-orbit coupling (SOC), however, are limited to the nodal line connections with the permanent shapes, leaving possible transformations between the different nodal net textures unexplored. Materials with the tunable nodal line connections would allow engineering the desired spinfull fermionic properties of the nodal line semimetals, thus providing a promising platform to discover new physics and to design potential applications.

In this work, we propose a feasible approach to realize tunable nodal line connections in real materials with non-negligible SOC. We identify space group symmetries that support the coexistence of symmetry enforced and accidental nodal lines. The former are robust to perturbations, while the latter are symmetry protected but tunable, i.e. can be created, altered, and annihilated by external stimuli. Combining the two types of the nodal lines in a single compound allows the tunability of the nodal nets by chemical substitution, pressure, or strain. Using density functional theory (DFT) calculations, we identify non-symmorphic quasi-one-dimensional (quasi-1D) compounds, $K_2SnBi$ and $MX_3$ (M = Ti, Zr, Hf and X = Cl, Br, I), as materials hosting the two-dimensional (2D) Dirac nodal nets that are robust to SOC. We show that in these compounds, the 2D Dirac nodal nets can be created and weaved into intricate patterns, using the symmetry enforced nodal lines as a frame and tuning the size and shape of the accidental nodal lines. Semimetals hosting such tunable 2D Dirac nodal nets are expected to reveal magneto-transport properties different from those previously known, due to the unique Landau levels controlled by the nodal net shape.

To realize the tunable 2D Dirac nodal nets we consider materials with a quasi-1D chain-like structure, which supports the inversion symmetry $P$ and the time reversal symmetry $T$, and has an off-centered mirror symmetry plane $g_i = \{M_i | (t_\parallel = 0; t_\perp \neq 0)\}$ perpendicular to the chain direction and a glide (or mirror) symmetry plane $g_j = \{M_j | (t_\parallel; t_\perp)\}$ parallel to the chain direction, such that $g_i$ and $g_j$ anticommute, i.e. $\{g_i, g_j\} = 0$. Here $M_i$ ($M_j$) is a mirror reflection about the plane perpendicular to the Cartesian axis $i$ ($j$); and $t_\parallel$ and $t_\perp$ are translations parallel and perpendicular to the mirror plane, respectively. As shown below, these conditions are sufficient to support the coexistence of the symmetry enforced and accidental nodal lines. The quasi-1D nature of these materials is required to efficiently tune their electronic structure through the interchain coupling, which is sensitive to external perturbations and the ionic radii of the chemical constituents.

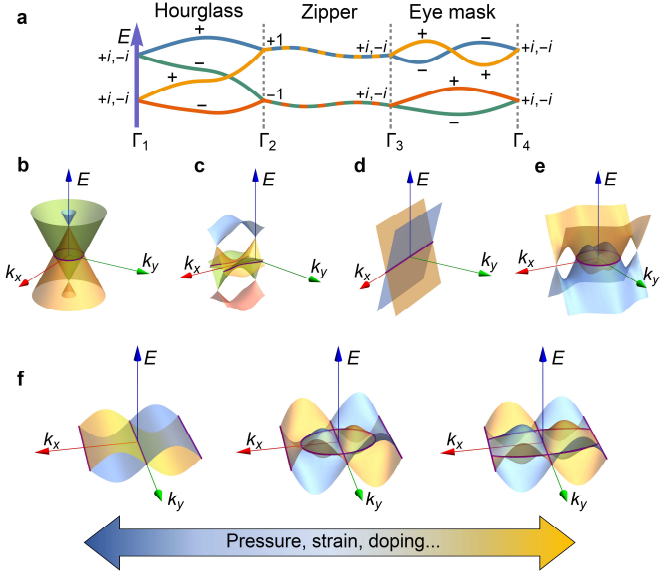

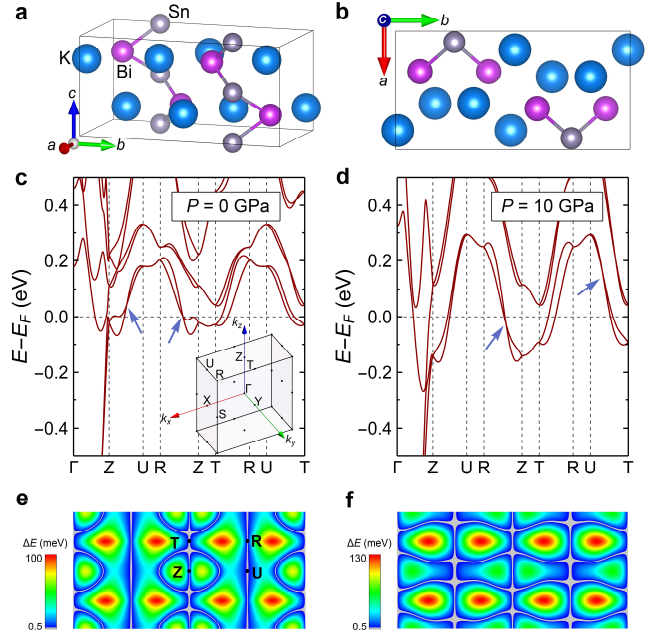

**Figure 1:** (a) Schematic of "double hourglass", "double zipper", and "double eye mask" band connectivities between the time reversal invariant momenta ($\Gamma_i$) in a glide invariant plane. Glide eigenvalues $\mu_i$ are indicated for the $\Gamma_i$ points and bands. (b, c) A symmetry enforced closed Dirac nodal loop (b) and a symmetry enforced open Dirac nodal line (c) from the double hourglass connectivity between $\Gamma_1$ to $\Gamma_2$. (d) A symmetry enforced open straight Dirac nodal line from the double zipper connectivity between $\Gamma_2$ to $\Gamma_3$. (e) A symmetry protected closed Dirac nodal loop from the accidental double eye mask connectivity. (f) Combination of the double zipper and double eye-mask Dirac nodal lines forming tunable 2D Dirac nodal nets.

**Figure 2**: (a) Crystal structure of $K_2SnBi$. (b) The structure viewed along the [001] axis. (c,d) Band structures of pristine $K_2SnBi$ (c) and $K_2SnBi$ under isostatic pressure of 10 GPa (d). The inset in (c) shows the Brillouin zone. Blue arrows point to the accidental eye mask band crossings. (e,f) Color maps of the band gap between the two doubly degenerate bands ($\Delta E$) in the $k_z = \frac{\pi}{c}$ plane for pristine $K_2SnBi$ (e) and $K_2SnBi$ under an isostatic pressure of 10 GPa (f). Grey contrast reflects the nodal net patterns (a band gap smaller than 0.5 meV).

A Dirac nodal line robust to SOC is protected by the combination of the symmetry operations $P$, $T$, and $g_i$ (see Supplemental Material [42] and refs. [20, 43, 44]). There are two types of the symmetry-enforced Dirac nodal lines. The first type is due to "double hourglass" band connectivity shown schematically in Fig. 1(a) between the $\Gamma_1$ and $\Gamma_2$ points [45]. This kind of connectivity is guaranteed by the $g_i = \{M_i|(t_\parallel \neq 0; t_\perp \neq 0)\}$ symmetry, which enforces a closed (loop-like) nodal line (Fig. 1(b)) or an open nodal line (Fig. 1(c)) in generic $k$-points [29, 31, 43, 44]. The second type is due to "double zipper" band connectivity shown in Fig. 1(a) between the $\Gamma_2$ and $\Gamma_3$ points. This kind of connectivity requires an additional perpendicular glide (or mirror) plane $g_j$, such that $\{g_i, g_j\} = 0$. In this case, along certain high symmetry paths, the two doublets are enforced to be locked to each other like a zipper, forming four-fold degenerate bands [42]. This double zipper connectivity enforces a straight nodal line along that path (Fig. 1(d)) [26, 46].

In addition to the symmetry enforced nodal lines, there exist symmetry protected nodal lines resulting from accidental degeneracy [44, 47]. Here we focus on the nodal lines formed by "double eye mask" connectivity shown schematically in Fig. 1(a) between the $\Gamma_3$ and $\Gamma_4$ points. This type of connectivity is protected by the $g_i = \{M_i|(t_\parallel = 0; t_\perp \neq 0)\}$ symmetry [42].

Two intersecting Dirac nodal lines must have the same electron filling (i.e., the number of occupied bands below the band crossing). From the simple electron counting, the double zipper and the double eye mask nodal lines have the same filling number. Therefore, a tunable 2D Dirac nodal net robust to SOC can be engineered by tailoring the size and shape of the double eye mask accidental nodal lines in the frame of double zipper nodal lines, as shown in Fig. 1(f). Suitable space groups to realize the nodal net tunability are summarized in Table S3 [42]. Below we demonstrate the feasibility of our predictions for real materials.

First, we consider $K_2SnBi$. The crystal structure of $K_2SnBi$ is characterized by two quasi-1D zigzag Sn-Bi chains along the $z$ direction separated by the intercalated K atoms in its simple orthorhombic unit cell (Figs. 2(a) and 2(b)) [48]. This compound belongs to the non-symmorphic space group 57 (Pbcm) [48], which has $P$, $g_x = \{M_x|(0, \frac{1}{2}, 0)\}$, $g_y = \{M_y|(0, \frac{1}{2}, \frac{1}{2})\}$, and $g_z = \{M_z|(0, 0, \frac{1}{2})\}$ symmetries. There are two parallel double zipper nodal lines along the high-symmetry U-R and Z-T lines in the Brillouin zone (inset in Fig. 2 (c)) enforced by the anticommutation relation $\{g_z, g_x\} = 0$ [42].

Figure 2(c) shows the calculated band structure of the pristine K$_2$SnBi [42]. Strong dispersions are seen along the Γ–Z direction due to the quasi-1D nature of the Sn-Bi chains. We focus on the two less dispersive doubly degenerate bands in the $k_z = \frac{\pi}{c}$ plane. Fig. 2(e) shows the calculated energy gap between these bands in the $k_z = \frac{\pi}{c}$ plane, which reflects the resulting nodal line pattern. The double zipper connectivities along the U-R and Z-T lines (Fig. 2(c)) enforce two parallel Dirac nodal lines along the U-R and Z-T directions (Fig. 2(e)). There are also the double eye mask band crossings in the Z-U and Z-R directions (Fig. 2(c)). The resulting ellipse-like accidental Dirac nodal line surrounding the Z point crosses one of the double zipper nodal lines (Fig. 2(e)).

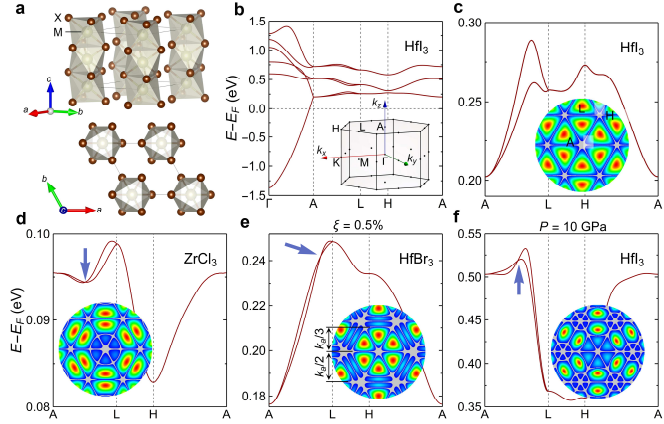

**Figure 3:** (a) The side view and top view of crystal structure of MX$_3$. (b) Band structure of pristine HfI$_3$. The inset shows the Brillouin zone. (c) Zoomed-in bands close to $E_F$ along the high symmetry lines in the $k_z = \frac{\pi}{c}$ plane and the related gap color map for HfI$_3$. (d)-(f) Band structures and gap color maps in the $k_z = \frac{\pi}{c}$ plane calculated for the pristine ZrCl$_3$ (d), HfBr$_3$ under a tensile strain of 0.5% (e), and HfI$_3$ under an isostatic pressure of 10 GPa (f). Blue arrows point to the accidental double eye mask band crossing. Grey contrast reflects the nodal net patterns (a band gap smaller than 0.1 meV).

These features of the 2D Dirac nodal net can be tuned by pressure, affecting the size and the shape of the accidental nodal lines. Figure 2(d) shows the calculated band structure of K$_2$SnBi under isostatic pressure of 10 GPa. The double eye mask band crossing in the Z-U direction is eliminated, while a new double eye mask crossing emerges in the U-T direction. This enlarges the ellipse-like accidental Dirac nodal line and links it to the double zipper Dirac nodal line along the U-R direction, forming a 2D Dirac nodal net over the whole $k_z = \frac{\pi}{c}$ plane (Fig. 2(f)). We find that a significant portion of the 2D Dirac nodal net in K$_2$SnBi emerges around the Fermi energy, which makes this material promising for experimental verification.

Next, we consider the Dirac nodal nets emerging in MX$_3$ compounds (M = Ti, Zr, Hf and X = Cl, Br, I) [49-51]. Figure 3(a) shows the atomic structure of MX$_3$, where M atoms are surrounded by octahedra of X atoms, which share common surfaces forming linear chains along the $z$ direction. These chains are arranged into a triangular lattice in the *x-y* plane. MX$_3$ belongs to the hexagonal space group 193 (P6$_3$/mcm), which has $P$, $g_z = \{M_z|(0,0,\frac{1}{2})\}$, $M_y$, and $C_{3z}$ (three-fold rotation about the *z* axis) symmetries. The anticommutation relation $\{g_z, M_y\} = 0$ enforces the double zipper Dirac nodal line to appear at $k_y = \frac{2n\pi}{\sqrt{3}a}$ ($n$ = integer). The $C_{3z}$ rotation applied to such Dirac nodal lines can generate other Dirac nodal lines at $k_y = \pm\sqrt{3}k_x + \frac{4n\pi}{\sqrt{3}a}$ [42].

Figures 3(b-f) show the calculated band structures of a series of MX$_3$ compounds. We focus first on HfI$_3$ as a representative material, where a possibility of the Dirac fermion has been discussed [52]. As seen from Fig. 3(b), the highest occupied band, largely composed of the Hf $d_{z^2}$ orbital [42], exhibits a strong dispersion along the Γ-A direction due to the quasi-1D crystal structure. There are two less dispersive doubly degenerate bands in the $k_z = \frac{\pi}{c}$ plane slightly above $E_F$, which are enlarged in Fig. 3(c). As expected from the symmetry, the double zipper bands appear along the L-H-A direction.

Chemical substitution in the MX$_3$ compounds allows introducing accident nodal loops into the nodal net pattern. We find that the double eye mask band crossings emerge in pristine ZrCl$_3$ (Fig. 3(d)), ZrBr$_3$, HfBr$_3$, and HfCl$_3$ (Fig. S8 [42]). These crossings produce accidental Dirac nodal loops centered around the A point. The accidental nodal line is circular in MCl$_3$ (inset in Fig. 3(d)) and hexagonal in MBr$_3$ (Fig. S8). Combination of such accidental nodal lines and the symmetry enforced nodal lines produces a 2D Dirac nodal net.

The size and shape of the accidental nodal loops can be further tuned by strain and pressure. For example, the outer symmetry enforced and inner accidental nodal hexagons in HfBr$_3$ form a hexagonal "spiderweb" pattern (Fig. S8). With in-plane tensile strain of 0.5%, the side-to-center distance of the inner hexagons can be tuned to $k_a$/3 (Fig. 3(e)). In this case, along the $k_y$ direction, the band crossings appear at "quantized" momenta $k_y = \pm k_a, \pm k_a/2, \pm k_a/3$. Moreover, using pressure one can not only turn on the accidental loop surrounding the A point in HfI$_3$, but also introduce an additional nodal loop surrounding the H point (Fig. 3(f)).

In order to demonstrate the uniqueness of such tunable 2D Dirac nodal nets, we analyze the Landau level spectra, which play a key role in magneto-transport properties. We derive the effective Hamiltonian from the tight-binding model for MX$_3$ near the L point $(0, \frac{2\pi}{\sqrt{3}a}, \frac{\pi}{c})$ in the Brillouin zone: $H(q) =$

$\alpha q_z \tau_x - q_y(\lambda_1 - \lambda_2 q_x^2 - \lambda_3 q_y^2)\tau_y \sigma_z$, where $q = k - k_L$, and $\tau$ and $\sigma$ are the Pauli matrices acting on the sublattice and spin degrees of freedom, respectively [42]. When $q_z = 0$, band degeneracy occurs at $q_y = 0$ (symmetry enforced) and at

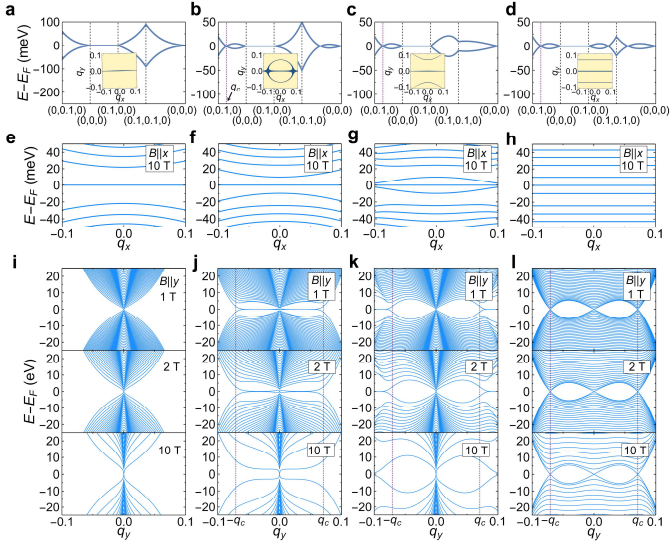

**Figure 4:** (a-d) Band structures calculated using the effective Hamiltonian for a symmetry enforced straight nodal line (a); a symmetry enforced straight nodal line plus an accidental ellipse-like nodal loop (b); a symmetry enforced straight nodal line plus accidental hyperbola-like nodal lines (c); a symmetry enforced straight nodal line plus accidental straight nodal lines (d). Insets show the nodal line patterns. (e-h) Landau level spectra for the nodal lines in (a-d) respectively for a magnetic field along the *x*-direction. (i-l) Landau level spectra for a magnetic field along the *y*-direction. The purple dashed lines denote the intersection of the accidental nodal line and $q_y$ axis in $q_y = \pm q_c$.

$q_y^2 = (\lambda_1 - \lambda_2 q_x^2)/\lambda_3$ (accidental). If $\lambda_2$ and $\lambda_3$ have the same sign, while $\lambda_1$ has an opposite sign, only a nodal line along the $q_x$ axis is enforced by symmetry (Fig. 4(a)). In other cases, accidental nodal lines can be ellipse-like, hyperbola-like, or straight parallel lines (Figs. 4(b-d) and Fig. S11). The nodal lines patterns, resulting from the effective Hamiltonian, capture the essential features of the 2D Dirac nodal nets calculated from first-principles [42].

Figs. 4(e-h) show the calculated Landau spectra under an in-plane magnetic field along the *x* direction (parallel to the symmetry enforced nodal line) for the nodal line patterns in Figs. 4(a-d). We find that the Landau levels are discrete with a four-fold degenerate flat band at zero energy. That is understandable since if we only keep the linear term in the continuum limit, the Hamiltonian resembles that of graphene [53]. A similar behavior has been suggested for a hypothetical hyper-honeycomb lattice with a Dirac nodal loop under a toroidal magnetic field [54]. Moreover, a flat band at $E_F$ implies the appearance of a pronounced peak in the electronic density of states, which can be detected experimentally by scanning tunneling spectroscopy, even if it is buried in other Fermi surfaces [27].

For an in-plane magnetic field along the *y* direction (perpendicular to the symmetry enforced nodal line), the Landau spectrum is gapless (Fig. 4(i)) when there is only one symmetry enforced nodal line along the $q_x$ axis (Fig. 4(a)). In this case, each Landau level is doubly degenerate. A crossing appears at $q_y = 0$, which consists of two pairs of counter-propagating chiral bands corresponding to the lowest Landau levels. This crossing is robust against the magnetic field and the emergent accidental nodal lines (Figs. 4(i-l)). In contrast to the known Dirac semimetals with a doubly degenerate crossing of the lowest Landau levels [2], the predicted crossing in the 2D Dirac nodal net compound is four-fold degenerate protected by symmetry [42].

Accidental nodal lines significantly modify the Landau spectra. In case of the ellipse-like accidental nodal loop (Fig. 4(b)) at a low magnetic field, a flat band with nearly four-fold degeneracy appears in a range of $0 < |q_y| < q_c$, i.e., between the edges of the symmetry enforced nodal line ($q_y = 0$) and the accidental nodal line (which intersects the $q_y$ axis at $q_y = \pm q_c$), as shown in Fig. 4(j). This result is consistent with the previous finding for a single accidental nodal loop in the presence of an in-plane magnetic field [27]. An increasing magnetic field splits and shrinks the flat band, isolating the symmetry-enforced single-crossing point (Fig. 4(j)). Therefore, the transition between a flat band and a single-crossing point can be induced by sufficiently large magnetic field. Similarly, flat bands appear in case of hyperbola-like accidental lines, but they have different ranges, i.e., $|q_y| > q_c$. An increasing magnetic field pushes the flat bands away from $q_y = \pm q_c$ (Fig. 4(k)). Interestingly, the four-fold degenerate flat band disappears when the accidental nodal lines become parallel to the symmetry enforced nodal line, which occurs at the critical point between the ellipse-like and hyperbola-like accidental nodal lines (Fig. 4(l)). When the nodal-line crossing occurs at $q_y = 0$, two additional four-fold degenerate crossings appear in the Landau spectrum and are located exactly at $q_y = \pm q_c$ (Fig. 4(l)). The magnetic field cannot shift or gap them. Therefore, the shape of the 2D Dirac nodal nets can be engineered to produce transitions between different types of the Landau spectra, i.e. those with a single-crossing point, multiple-crossing points, and flat bands, which are expected to have different magneto-transport properties.

We note that quasi-1D compounds might experience the Peierls distortion or a charge-density wave, leading to a trivial band insulator [55] or a topological phase [56]. Usually, the compounds with the Peierls distortion reveal phonon instability [57]. For $K_2SnBi$ and $HfI_3$, we have calculated the phonon dispersions (Fig. S9) and found no unstable phonon mode, indicating that these materials are stable in respect to the Peierls distortion.

In conclusion, we have proposed the realization of tunable 2D Dirac nodal nets in quasi-1D non-symmorphic compounds, resulting from the interplay between the coplanar symmetry enforced and symmetry protected accidental Dirac nodal lines. Using DFT calculations, we have demonstrated that the appearance, shape, and size of the 2D Dirac nodal nets in $K_2SnBi$ and $MX_3$ compounds can be tuned by chemical substitution, pressure, or strain, leading to controllable physical properties, as is evident from the nontrivial Landau level spectra in these compounds. Our work offers a feasible plan to realize the tunability of the nodal line connections in real materials, which is important for exploring novel physical properties and potential applications of the nodal line semimetals.

This work was supported by the National Science Foundation (NSF) through the Nebraska MRSEC program (grant DMR-1420645) and the DMREF program (grant DMR-1629270). S.-H.Z thanks the support of National Science Foundation of China (NSFC Grants No. 11504018). Computations were performed at the University of Nebraska Holland Computing Center. Figures were plotted using VESTA [58] and the SciDraw scientific figure preparation system [59].

* dfshao@unl.edu
† tsymbal@unl.edu
‡ These authors contributed equally to this work.